

\documentstyle{amsppt}
\NoBlackBoxes
\hoffset=.1in

\define\id{\operatorname{id}}

\define\ld{\ldots}

\define\a{\alpha}
\define\be{\beta}

\define\ve{\varepsilon}
\define\gm{\gamma}
\define\de{\delta}

\define\vd{\varDelta}
\define\vl{\varLambda}

\define\dert#1{\frac{#1}{dt}}
\define\pd#1#2{\frac{\partial#1}{\partial#2}}

\define\vc#1{(#1_1,\ldots,#1_n)}
\define\vct#1{[#1_1,\ldots,#1_n]}
\define\vect#1{\{#1_1,\ldots,#1_n\}}

\define\suk{\sum^n_{k=1}}


\define\prodn{\prod^n_{k=1}}

\define\jak{\dfrac{i}{\kappa}}
\define\xak{\dfrac{1}{\kappa}}


\define\frajx{\frac{1}{2}} 



\define\lra{\Leftrightarrow }
\define\lrar{\Longleftrightarrow }

\define\1{Poincar\'e group}
\define\2{Poincar\'e algebra}
\define\3{quantum}
\define\4{group}
\define\5{commut}
\define\6{representation}
\define\7{character}
\define\8{product}
\define\9{general}
\define\0{operation}

\define\irt{ \vartriangleright\!\blacktriangleleft }

\define\trir{ \triangleright }
\define\tril{ \triangleleft }

\topmatter
\title The duality between $\kappa$-Poincar\'e
algebra and $\kappa$-Poincar\'e  group
\endtitle
\rightheadtext{$\kappa$-Poincar\'e algebra and $\kappa$-Poincar\'e  group}
\leftheadtext{P. Kosi\'nski, P. Ma\'slanka}
\author  Piotr Kosi\'nski\\
{\it Department of Theoretical Physics}\\
{\it  University of \L \'od\'z}\\
{\it ul. Pomorska 149/153, 90--236 \L \'od\'z, Poland}\\
Pawe\l \/ Ma\'slanka\\
{\it Institute of Mathematics}\\
{\it  University of \L \'od\'z}\\
{\it ul. St. Banacha 22, 90--238, \L \'od\'z, Poland}
\endauthor
\thanks
*\ Supported by KBN grant 2P\,30\,2217\,06\,p\,02
\endthanks

\abstract The full duality between  the $\kappa$-Poincar\'e  algebra  and
$\kappa$-Poincar\'e group is proved.
\endabstract
\endtopmatter

\document
\head 1. Introduction
\endhead

Recently, much attention has been paid to the specific deformation of
Poincar\'e algebra --- the  so-called $\kappa$-Poincar\'e algebra [1]. Its
global counterpart,  the $\kappa$-\1, has been obtained by  S. Zakrzewski
[2]. His method consisted in quantising the Poisson structure on classical
$r$-matrix obtained from $\frac{1}{\kappa}$ expansion of algebra coproduct.
Zakrzewski method gives, in principle, the duality group $\lra$ algebra only
in the lowest, $\frac{1}{\kappa}$-approximation. However, due to the lack of
ordering ambiguities in the quantisation procedure it seemed likely there
is a full duality between $\kappa$-\1 and algebra. Indeed, it was shown ([3].
[4], [5]) that this is the case in two dimensions. The proof given in the
last paper relies heavily on the bicross\8 structure of $\kappa$-\2 and \4
discovered in [5] and [6]. Here,  we briefly sketch how to extend this proof to
four dimensions. In fact, it is easily seen that the proof presented below
works in any dimensions. The full version will appear elsewhere.

\head II. The $\kappa$-\2 and $\kappa$-\1
\endhead

The $\kappa$-\2 $\Cal P_k$ is defined by the following rules ([1], [6])
($M_i = \frajx \ve_{ijk} M_{jk}$, $N_i = M_{i0}$):
$$
\alignat 2
&[P_\mu,P_\nu] = 0, & \quad & \\
&[M_i,M_j] = i\ve_{ijk} M_k,     &\quad \quad \quad &   [M_i,P_0] =0,\\
&[M_i,N_j] = i\ve_{ijk} N_k,     &\quad \quad \quad \quad \quad \quad  &
[M_i,P_j] = i\ve_{ijk} P_k,\\
&[N_i,N_j] = -i\ve_{ijk} M_k,     &\quad \quad \quad \quad \quad \quad &
[N_i,P_0] = iP_i,
\endalignat
$$
$$
[N_i,P_j] = i\de_{ij} \Bigg(\dfrac{\kappa}{2}\Big( 1 -
e^{-\frac{2P_0}{\kappa}} \Big) + \dfrac{1}{2\kappa} \vec{P\,}^2 \Bigg) -
\dfrac{i}{\kappa} P_iP_j,
\tag{1}
$$
$$
\align
& \vd P_0 = P_0  \otimes  I + I   \otimes P_0,   \qquad\qquad  \vd P_i = P_i
\otimes  e^{-\frac{P_0}{\kappa}}  + I  \otimes P_i,   \\
& M_i = M_i  \otimes  I + I  \otimes  M_i,  \\
& \vd N_i = N_i    \otimes  e^{-\frac{P_0}{\kappa}} + I  \otimes N_i - \xak
\ve_{ijk}  M_j  \otimes P_k,   \\
& S(P_\mu) = -P_\mu, \qquad S(M_i) = -M_i,  \qquad   S(N_i) = -N_i +
\dfrac{3i}{2\kappa}  P_i,\\
& \ve(X) = 0,  \qquad\qquad X = P_\mu, M_i,N_i .
\endalign
$$
It has ([5], [6]) a natural bicross\8 ([7]) structure
$$
\Cal P_k = T  \blacktriangleright\!\vartriangleleft U(s0(3,1)).
\tag{2}
$$
Indeed, it is sufficient to define
$$
\alignat 2
& M_i \trir P_0 = 0,    &\quad &   M_i \trir P_k  = i\ve_{ijk} P_j,\\
& N_i \trir P_0 = iP_i ,      &\quad &   N_i\trir P_j = i \de_{ij}
 \Bigg(\dfrac{\kappa}{2}\Big( 1 -
e^{-\frac{2P_0}{\kappa}} \Big) + \dfrac{1}{2\kappa} \vec{P\,}^2 \Bigg) -
\dfrac{i}{\kappa} P_iP_j, \tag{3}\\
& \de(M_i) = M_i   \otimes I,       &\quad &   \de(N_i) = N_i  \otimes
e^{-\frac{P_0}{\kappa}} - \xak \ve_{ijk} M_j \otimes  P_k
\endalignat
$$
here $(N_i,M_i)$ is the standard Lorentz algebra (with standard co\8) while
$T$ is defined as an algebra generated by $P_\mu$, $\mu = 0,\ld,3$ obeying
the following relations:
$$
\alignat 2
& [P_\mu, P_\nu] = 0, &\quad & \\
& \vd P_0 = P_0 \otimes I + I \otimes  P_0,   &\quad &  \vd P_i = P_i
\otimes  e^{-\frac{P_0}{\kappa}} +   I \otimes  P_i,
 S(P_\mu) = - P_\mu,  \tag{4}\\
&   \ve(P_\mu) = 0. &\quad &
\endalignat
$$

{\bf II.2. The \ } $\kappa${\bf -Poincar\'e group.\ } The $\kappa$-\1
$\widetilde{\Cal P}_\kappa$ is defined by the following relations ([2],
[5]):
$$
\aligned
& [x^\mu,x^\nu  ] = \jak (\de^\mu_0 x^\nu - \de^\nu_0 x^\mu),    \\
& [\vl^\mu{}_\nu, x^\rho] =  - \jak ((\vl^\mu{}_0 - \de^\mu_0)
\vl^\rho{}_\nu + (\vl^0{}_\nu - \de^0_\nu)q^{\mu\rho}),  \\
&       [\vl^\mu{}_\nu, \vl^\a{}_\be] = 0, \\
& \vd( \vl^\mu{}_\nu) = \vl^\mu{}_\a \otimes   \vl^\a{}_\nu,    \quad
\quad  \vd(x^\mu ) =   \vl^\mu{}_\a \otimes x^\a +   x^\mu \otimes I, \\
&       S(\vl^\mu{}_\nu)  = \vl_\nu{}^\mu,    \quad \quad \quad   \quad  \quad
  S(
x^\mu) = -   \vl_\nu{}^\mu x^\nu, \\
&  \ve( x^\mu   ) = 0, \quad  \quad  \quad    \quad  \quad  \quad  \quad
\ve(\vl^\mu{}_\nu) = \de^\mu_\nu.
\endaligned
\tag{5}
$$
Again, $\widetilde{\Cal P}_\kappa$ can be defined as a bicross\8 ([5]):
$$
\widetilde{\Cal P}_\kappa = T^* \irt C(S0(3,1)).
$$
To see this it is sufficient to define ([5]):
$$
\be(x^\mu) = \vl^\mu{}_\nu \otimes x^\nu.
$$
$$
\vl^\mu{}_\nu \tril x^\rho = - \jak((\vl^\mu{}_0 - \de^\mu_0)
\vl^\rho{}_\nu + (\vl^0{}_\nu - \de^0_\nu)g^{\mu\rho}) = [\vl^\mu{}_\nu,
x^\rho].
$$
Moreover, while $ C(S0(3,1))$ is the standard algebra of functions defined
over Lorentz group while $T^*$ is defined by the following relations:
$$
[x^\mu,x^\nu] = \jak (\de^\mu_0 x^\nu - \de^\nu_0 x^\mu),
$$
$$
\vd(x^\mu) =  x^\mu  \otimes I + I  \otimes x^\mu,
$$
$$
S(x^\mu) = -   x^\mu, \qquad  \ve(x^\mu) = 0.
$$

{\bf II.3. Duality. \ } We shall define the dualities:
$$
C(S0(3,1))  \lrar   U(s0(3,1)),
$$
$$
T^* \lrar  T.
$$
First, we define standard duality between the  Lorentz group and algebra as
follows:
$$
\aligned
\langle \vl^\mu{}_\nu, M_{\a\be} \rangle    & \equiv i \dert d (
e^{itM_{\a\be}})^\mu{}_\nu\Big|_{t = 0} = (M_{\a\be})^\mu_\nu\\
& = i( \de^\mu_\a g_{\nu\be} - \de^\mu_\be   g_{\nu\a}).
\endaligned
\tag{6}
$$

The following lemma is obvious.

\proclaim{Lemma 1}
$$
\langle \vl^{\mu_1}{}_{\nu_1} \cdots  \vl^{\mu_n}{}_{\nu_n}, M_{\a\be}
\rangle = i \suk ( \de^{\mu_k}_\a g_{\nu_k\be} - \de^{\mu_k}_\be
g_{\nu_k\a}) \prod_{l\ne k} \de^{\mu_l}_{\nu_l}.
\tag{7}
$$
\endproclaim

The duality $T^* \lra T$ is defined by
$$
\langle x^\mu,p_\nu \rangle = i\de^{\mu}_{\nu}.
$$
This duality can be fully described as follows. For any function
$\psi(x^\mu)$ we define the normal \8  $:\psi(x^\mu):$ as the one in which all
$x^0$ factors stand leftmost. We  then have

\proclaim{Lemma 2 {\rm ([6])}}
$$
\langle : F(x^\mu):, f(p_\nu) \rangle = f\Bigg(i \pd {}{x^\nu}\Bigg)
F(x^{\mu}) \Big|_{x=0}.
\tag{8}
$$
\endproclaim

A simple proof is based on Leibnitz rule and the identity
$$
(x^n)^l (x^0)^n = \Big(x^0 - i \dfrac{l}{\kappa}\Big)^n(x^m)^l.
\tag{9}
$$

{\bf II.3. The structure of $\tril$ and $\be$ operations. \ } In the sequel we
shall need some more detailed information concerning the structure of the
operations $\tril$ and $\be$. We have the following lemma.

\proclaim{Lemma 3}
$$
(\vl^{\mu_1}{}_{\nu_1} \cdots  \vl^{\mu_n}{}_{\nu_n}) \tril x^{\rho_1}
\tril  \cdots    \tril   x^{\rho_m} = [  \cdots  [ \vl^{\mu_1}{}_{\nu_1}
\cdots  \vl^{\mu_n}{}_{\nu_n},\, x^{\rho_1}],  \ldots  , x^{\rho_m}].
\tag{10}
$$
\endproclaim

The proof is based on the following rule:
$$
(ab) \tril h = (a  \tril h_{(1)}) (b  \tril h_{(2)}).
\tag{11}
$$
It is slightly more difficult to describe the structure of $\be$ \0. To this
end we define the $\tau$-\0, acting on an arbitrary \8 of $\vl$'s and $x$'s
as follows: using commutation rules (5), we transpose all $x$'s to the left
and then put $x^\mu = 0$. By linearity we extend $\tau$ to any polynomial in
$x$'s and  $\vl$'s. The $\tau$-\0 has the following obvious property
$$
\tau ( \tau(P_1)P_2) = \tau (P_1P_2).
\tag{12}
$$

\proclaim{Lemma 4}
$$
\be(x^{\mu_1} \cdots  x^{\mu_n}) = (\tau \otimes \id) \Bigg(\prodn (
\vl^{\mu_k}{}_{\nu_k} \otimes x^{\nu_k} + x^{\mu_k} \otimes I)\Bigg).
\tag{13}
$$
\endproclaim

The inductive proof is based on identity (12) and the product rule ([5],
[6], [7]):
$$
\be (hg) = (h^{\bar 1}  \tril g_{(1)})   g_{(2)}^{\bar 1} \otimes h^{\bar
2}   g_{(2)}^{\bar 2} .
\tag{14}
$$

\head III. The proof of duality
\endhead

We have to prove the following duality relations:
$$
\langle X,  M_{\a\be}  \trir  P_\gm \rangle =
\langle \be(X),  M_{\a\be}  \otimes P_\gm \rangle,
\tag{15a}
$$
$$
\langle \vl \tril X,  M_{\a\be}\rangle =
\langle \vl  \otimes X,  \de(M_{\a\be})\rangle,
\tag{15b}
$$
here $X$ is an arbitrary \8 of  $x$'s while $\vl$ is   an arbitrary \8 of
$\vl$'s. We assume that the \0s $\tril$ and $\be$ are known and use
relations (15) to prove the structure of $\trir$ and $\de$

\proclaim{Theorem 1} The following rules are implied by (15a)
$$
\aligned
& M_i \trir P_0 = 0, \qquad  M_i\trir P_j = i \ve_{ijk}P_k, \qquad N_i\trir
P_0 = i P_i,\\
& N_i \trir P_j = i\de_{ij} \Bigg(\dfrac{\kappa}{2}\Big( 1 -
e^{-\frac{2P_0}{\kappa}} \Big) + \dfrac{1}{2\kappa} \vec{P\,}^2 \Bigg) -
   \dfrac{i}{\kappa} P_iP_j.
\endaligned
\tag{16}
$$
\endproclaim

As an example we shall prove the most complicated last equality. It follows
immediately from (8) and the following lemma

\proclaim{Lemma 5}
$$
\langle \be(x^{m_1} \cdots  x^{m_r}), M_{i0}  \otimes P_k \rangle =
\cases
0, &\qquad r \ne 2,\\
    \jak(\de_{m_1k} \de_{m_2i} + \de_{m_1i} \de_{m_2k} - \de_{m_1m_2}
\de_{ik}), & \qquad r = 2.
\endcases
\tag"{(i)}"
$$
$$
\langle \be((x^0)^n x^{m_1} \cdots  x^{m_r}), M_{i0}  \otimes P_k \rangle =
\cases
0, &\qquad r  > 0,\\
    - \de_{ik} \Bigg( - \dfrac{2i}{\kappa}\Bigg)^{n-1},
 & \qquad r = 0.
\endcases
\tag"{(ii)}"
$$
\endproclaim

\demo{Proof} We prove, for example, the first part of the lemma. First, we
have
$$
\align
\langle \be(x^m),M_{i0}  \otimes P_k \rangle & =  \langle \vl^m{}_\nu
\otimes  x^\nu,M_{i0}  \otimes P_k \rangle\\
& = i  \langle \vl^m{}_k , M_{i0}  \rangle
\endalign
$$
and by Lemma 4
$$
\align
\langle \be(x^m x^n),M_{i0}  \otimes P_k \rangle & =  \langle \vl^m{}_\mu
\vl^n{}_\nu \otimes x^\mu x^\nu,M_{i0}  \otimes P_k \rangle\\
& + \langle [\vl^m{}_\mu, x^n] \otimes x^\mu,M_{i0}  \otimes P_k \rangle  .
\endalign
$$
The second term on the right-hand side can be evaluated immediately using
Lemmas 1 and 2. In order to evaluate the first one let us notice that, by
Lemma 2 and the commutation rule for $x$'s the only term that gives a
nonvanishing contribution corresponds to $\mu = k$, $\nu = 0$. Let us now
consider the case $r > 2$. First, note that
$$
\langle x^{\mu_1} \cdots  x^{\mu_n},P_k \rangle  = \Big( -\jak\Big)^{n-1}
\de_{\mu_1k}\prod^n_{l=2}  \de_{\mu_l0}.
\tag{17}
$$
Therefore, we have by (17) and the definition of $\tau$
$$
\aligned
\langle \be(x^{m_1} & \cdots  x^{m_r} ),M_{i0}  \otimes P_k \rangle\\
&= \Big\langle (\tau \otimes \id) \Big(\prod^r_{l=1}  (
\vl^{m_l}{}_{\nu_l} \otimes x^{\nu_l} + x^{m_l} \otimes I)\Big),
M_{i0}  \otimes  P_k  \Big\rangle\\
&= \Big\langle (\tau \otimes \id) \Big((\vl^{m_1}{}_k  \otimes x^k)
\prod^r_{l=2}  (\vl^{m_l}{}_0  \otimes x^0  + x^{m_l} \otimes
I)\Big), M_{i0}   \otimes P_k \Big\rangle.
\endaligned
\tag{18}
$$
We shall prove that for $r \ge 3$
$$
 (\tau \otimes \id)  \Bigg((\vl^{m_1}{}_k  \otimes x^k)
\prod^r_{l=2}  (\vl^{m_l}{}_0  \otimes x^0  + x^{m_l} \otimes
I)\Bigg) = \vl^{m_1 \cdots m_r}_A \otimes x^A
\tag{19}
$$
where, for any multiindex $A$, $\vl^{m_1 \cdots m_r}_A$ can be decomposed
into the sum of monomials, each containing $\vl^0_0 - 1$ or$\slash$and
$\vl^m{}_0 \vl^n{}_0$ or$\slash$and   $\vl^0{}_m \vl^0{}_n$ or$\slash$and
$\vl^0{}_m \vl^n{}_0$

In order to prove this we use induction with respect to $r$. We have by (12)
$$
\align
(\tau & \otimes \id) \Big((\vl^{m_1}{}_k  \otimes x^k)
\prod^{r+1}_{l=2}  (\vl^{m_l}{}_0  \otimes x^0  + x^{m_l} \otimes
I)\Big) = (\tau \otimes \id) \\
& \cdot \Bigg(\Big((\tau \otimes \id)(\vl^{m_1}{}_k  \otimes x^k)
\prod^{r}_{l=2}  (\vl^{m_l}{}_0  \otimes x^0  + x^0 \otimes I)\Big)
(\vl^{m_{r+1}}_0    \otimes x^0  + x^{m_{r+1}} \otimes I) \Bigg) \\
& = (\tau \otimes \id) ((\vl^{m_1 \cdots m_r}_A \otimes
x^A)(\vl^{m_{r+1}}_0    \otimes x^0  + x^{m_{r+1}} \otimes I))\\
& =  \vl^{m_1 \cdots m_r}_A      \vl^{m_{r+1}}_0    \otimes x^A x^0 +
[\vl^{m_1 \cdots m_r}_A,x^{m_{r+1}}] \otimes x^A.
\endalign
  $$
The first term on the right-hand side has already the proper structure. In
order to prove the same for the second term it is sufficient to use the
following commutation rules:
$$
\aligned
& [\vl^0{}_0 - 1, x^n] = - \jak (\vl^0{}_0 - 1) \vl^n{}_0 ,\\
& [\vl^0{}_k, x^n] = - \jak (\vl^0{}_0 - 1) \vl^n{}_k ,\\
& [\vl^k{}_0    , x^n] = - \jak (\vl^n{}_0 \vl^k{}_0 + g^{nk}(\vl^0{}_0 -
1)) .\\
\endaligned
\tag{20}
$$
For $r = 3$ relation (15) is verified by simple straightforward calculation.

In order to complete the proof of Lemma (5i) we note that, by Lemma 1,
$$
\langle \vl, M_{\a\be} \rangle = 0
$$
if $\vl$ is any monomial containing $\vl^0{}_0 - 1$ or$\slash$and
$\vl^m{}_0 \vl^n{}_0$ or$\slash$and   $\vl^0{}_m \vl^0{}_n$ or$\slash$and
$\vl^0{}_m \vl^n{}_0$.

Relation (ii) can be proved along the same lines.
\enddemo
\proclaim{Theorem 2} The following rules are implied by (15b):
$$
\align
&\de(M_i) = M_i  \otimes  I,\\
&\de(N_i) = N_i  \otimes  e^{-\frac{P_0}{\kappa}} - \xak \ve_{ijk}  M_j
\otimes  P_k.
\endalign
$$
\endproclaim

Again, as an example, we prove the last equality. It follows from (8) and
the following

\proclaim{Lemma 6}
\roster
\item"{(i)}"
$$
\Big\langle \prod_a \vl^{\mu_a}{}_{\nu_a}   \tril (x^0)^n, M_{i0}\Big\rangle
= \Big( -\jak\Big)^n \Big\langle \prod_a \vl^{\mu_a}{}_{\nu_a}   , M_{i0}
\Big\rangle,
\tag{21}
$$
\item"{(ii)}" for $r > 0$
$$
\Big\langle \prod_a \vl^{\mu_a}{}_{\nu_a}   \tril (x^0)^l   \prod^r_{p=1}
x^{m_p}, M_{i0}\Big\rangle
= \Big( -\jak\Big)   \de_{l0}\de_{r1}  \Big\langle \prod_a
\vl^{\mu_a}{}_{\nu_a}   , M_{m_1i} \Big\rangle.
\tag{22}
$$
\endroster
\endproclaim

We shall prove (i). We use induction with respect to $n$. First, note the
following important property of the commutator $[\vl^\mu{}_\nu,x^\rho]$
$$
[\vl^\mu{}_\nu,x^\rho]\Big|_{\vl^\a{}_\be \to  \de^\a_\be} = 0.
\tag{23}
$$
Therefore, by Lemma 1, we have
$$
\Big\langle \prod_a \vl^{\mu_a}{}_{\nu_a}  \tril x^0 , M_{i0} \Big\rangle =
\Big\langle \Big[\prod_a \vl^{\mu_a}{}_{\nu_a},x^0\Big] , M_{i0}
\Big\rangle  = \sum_a \prod_{b\ne a}  \de^{\mu_b}_{\nu_b} \langle [
\vl^{\mu_a}{}_{\nu_a}, x^0 ], M_{i0} \rangle.
$$
But, by straightforward calculation
$$
\langle [ \vl^{\mu_a}{}_{\nu_a}, x^0 ], M_{i0} \rangle =  \Big( -\jak\Big)
 \langle \vl^{\mu_a}{}_{\nu_a}   , M_{i0}  \rangle
$$
so, appealing again to Lemma 1, we get equality (21) for $n = 1$.

For $n > 1$ we use again (23) and Lemma 3 to infer
$$
\Big\langle \prod_a  \vl^{\mu_a}{}_{\nu_a} \tril (x^0 )^n, M_{i0}
\Big\rangle  = \sum_a \prod_{b\ne a}  \de^{\mu_b}_{\nu_b} \langle
\vl^{\mu_a}{}_
{\nu_a}
\tril (x^0)^n, M_{i0} \rangle.
$$
By induction hypothesis
$$
\align
 \langle  \vl^{\mu_a}{}_{\nu_a} \tril (x^0)^{n+1}, M_{i0} \rangle
& =   \langle [\vl^{\mu_a}{}_{\nu_a} x^0] \tril (x^0)^{n}, M_{i0} \rangle \\
& = \Big( -\jak\Big)^n  \langle [ \vl^{\mu_a}{}_{\nu_a}, x^0 ], M_{i0}
\rangle = \Big( -\jak\Big)^{n+1}  \langle  \vl^{\mu_a}{}_{\nu_a}, M_{i0}
\rangle
\endalign
$$
which gives (i); (ii) can be proved in a similar way.

Now, the full duality follows from the general theory of bicross\8s.

\enddocument